# MR-Compass: Inertial Navigation–Driven Motion Correction for Brain MRI

Musa Tunc Arslan[1,2,*], Fatih Calakli[1,2], Joshua Auger[1,2], Hongli Fan[3,†], Alan J. Macy[4,††], and Simon K. Warfield[1,2]

[1]Computational Radiology Laboratory, Department of Radiology, Boston Children's Hospital, MA, US.
[2]Department of Radiology, Harvard Medical School, MA, US.
[3]Siemens Medical Solutions, Siemens, MA, US.
[4]Research and Development Director, BIOPAC Systems Inc., CA, US.
[*]Address correspondence to: musa.arslan@childrens.harvard.edu

**Abstract**

Inertial sensors can track object kinematics, however, unbounded drift from integrating noisy signals makes them impractical for MRI motion correction at millimeter resolution and minute-long scans. We introduce MR-Compass, which exploits the MRI system's static magnetic and gravitational fields to estimate 3-DOF orientation at 2 kHz directly, without integration, eliminating random-walk. The remaining 3-DOF translation is recovered via phase correlation from the MRI data. We experimentally validate the efficacy of the method retrospectively using a 3D radial koosh-ball sequence and prospectively using 2D EPI fMRI during large volunteer motions. MR-Compass followed by phase-correlation achieved a mean accuracy of $0.6^o$ and 0.4 pixels across all experiments. Image quality improved when motion correction was applied in all volunteer scans for both retrospective and prospective correction cases. MR-Compass was effective in measuring head motion in the MRI scanner with high accuracy at unprecedented sample rates, and enabled both retrospective and prospective reconstruction to improve image quality by aligning the k-space data appropriately and by reducing the motion related artifacts.

## 1   Introduction

MRI of the brain is one of the most common noninvasive diagnostic imaging procedures. MRI can provide critically important diagnostic information, but motion can degrade image quality. Motion artifacts compromise 14–35% of brain exams [1, 2] and are a particular challenge for pediatric, elderly, or uncooperative patients due to various reasons, such as neurodegenerative diseases (i.e., Parkinson's [3, 4]) or epilepsy [5, 6], often necessitating repeat scans. Preventive measures like physical restraints, training, sleep manipulation or anesthesia are not always practical [7–9].

---

Accordingly, three main strategies have been developed: retrospective motion correction (RMC), which modifies k-space data after acquisition [3]; prospective motion correction (PMC), which updates the scan in real time [10]; and hybrid approaches that combine both [11]. All rely on accurate motion estimation, where position and orientation are essential for compensating head motion in neuroimaging. Motion has been sensed using intrinsic MRI signals (images, ID, vNav, orbital navigators, guidance lines) [12–17], and external devices [13, 18–24].

Optical systems utilize markers (e.g., Kineticor [12]) or structured light (e.g., Tracoline [13, 18, 19]) to track either a marker object or the patient's face. However, both the marker and the face can be occluded by the head coil [11], preventing motion estimates, and can also yield erroneous estimates due to nonrigid skin motion such as facial gestures [17]. These errors degrade MRI images even in the absence of rigid motion. Furthermore, camera-based systems must align camera and MRI coordinate frames (hand-eye calibration), which is challenging and becomes invalid if the camera is moved between patients [19, 25]. Tracoline also requires a 3D facial mesh obtained from a separate, lengthy MRI scan [19, 26, 27]. Optical tracking performs well with cooperative volunteers but may fail with uncooperative subjects [18, 19, 27, 28].

Pilot tone–based motion tracking injects RF signals close to the NMR resonance frequency into the scanner room which are detected by all receiver coils [23, 24]. However, PT signals do not explicitly encode motion parameters, so a calibration step is required [29–31]. This calibration is performed by acquiring PT and MR data while patients hold static poses followed by image registration to derive the PT mapping [29, 30, 32]. Multiple poses are required, yet they may not span the full motion range; non-cooperative patients may be unable to comply, and the required calibration methods are patient specific, creating a workflow bottleneck [32].

MR-based motion tracking is often more convenient. Motion-robust k-space trajectories, such as radial lines [33], enable self-navigation through k-space center oversampling [15, 33–35], but increase reconstruction complexity and are restricted to specific sequences. Navigator methods acquire additional data for motion estimation and can be image-based (PROMO [16], vNav [14]), excitation-based (fatNav [36]), or k-space-based (spherical [37], orbital [38], cloverleaf [39]). A fundamental trade-off exists between navigator accuracy and acquisition time, limiting motion updates to once per repetition time (TR), often several seconds [14, 36, 37, 39]. Electromagnetic tracking systems [20, 21] manipulate MRI gradients to induce sensor signals, resulting in similar once per TR sampling rates.

In navigation, dead reckoning determines position using prior location, speed, heading, and time [40]. The position of an object can be tracked by double-integrating acceleration measured by an accelerometer circumventing the need for visual line-of-sight, provides high sampling rates and is not patient specific specific. However, accelerometers measure the sum of kinematic acceleration and orientation dependent gravity. To isolate the kinematic component, single integration of the angular velocity provides orientation[41, 42], and a magnetometer supplies directional reference. Together, these sensors form an Inertial Measurement Unit (IMU).

The core challenge is unbounded drift (random walk) resulting from the integration of sensor noise. This drift is more severe for position due to the double integration of acceleration. Orientation errors further corrupt kinematic estimates by leaking gravity into the measured acceleration. Various methods have been developed to mitigate this drift in the pedestrian navigation [43–46], achieving centimeters level accuracies.

MRI resolution is on the millimeter scale, with acquisitions lasting minutes. Consequently,

IMUs are rarely used for MRI motion correction. The only published example includes a phantom study limited to single axis rotations [47]. Rotations and translations about multiple axes were not addressed. Furthermore, van Niekerk et al. utilized a Madgwick filter that integrates angular velocity and is therefore still affected by the unbounded drift [47–49].

In this work, we introduce an "MR-Compass" for motion correction. Our MRI-compatible IMU records data at 2000 Hz. It utilizes the MRI's strong static magnetic and gravitational fields to compute the 3-DOF orientation directly, at high rate and without integration, thus avoiding random-walk drift. We show that the remaining 3-DOF translation can be estimated from the data via phase-correlation. We demonstrate the method's efficacy for motion correction through in-vivo experiments, using both a non-Cartesian T1-weighted 3D radial koosh ball (3D-radial) sequence retrospectively and a 2D Cartesian EPI fMRI sequence prospectively.

## 2 Theory

The aim of rigid motion correction is to determine the object's translation and rotation and apply them to the gradients, RF, and downconverters, ensuring that the k-space trajectory follows the object and the Bloch equations are satisfied. Assuming volumetric excitation, the MRI signal $s(t, \boldsymbol{G}(t))$ from a proton density $\rho(\boldsymbol{x})$ is as follows [50–55]:

$$s(t, \boldsymbol{G}(t)) = \int_V \rho(\boldsymbol{x}) \exp\left\{-i\gamma \int_0^t \boldsymbol{G}^\top(\tau)\boldsymbol{x} d\tau\right\} d\boldsymbol{x} \tag{1}$$

Here, $\boldsymbol{G}(t)$ are the applied gradients and superscript $\top$ is the transpose operation. If the object experiences a rigid body motion, the point $\boldsymbol{x}$ moves to a new position:

$$\boldsymbol{x}_n = \boldsymbol{R}\boldsymbol{x} + \boldsymbol{r} \tag{2}$$

where $\boldsymbol{R} \in \mathbb{R}^{3\times 3}$ is a rotation matrix and $\boldsymbol{r} \in \mathbb{R}^{3\times 1}$ is a translation vector. The MRI signal after the motion is as follows:

$$\hat{s}(t, \boldsymbol{G}(t)) = \int_V \rho(\boldsymbol{x}_n) \exp\left\{-i\gamma \int_0^t \boldsymbol{G}^\top(\tau)\boldsymbol{x} d\tau\right\} d\boldsymbol{x} \tag{3}$$

A change of variables $\boldsymbol{x} = \boldsymbol{R}^\top(\boldsymbol{x}_n - \boldsymbol{r})$ results in:

$$\hat{s}(t, \boldsymbol{G}(t)) = \left[\int_V \rho(\boldsymbol{x}_n) \exp\left\{-i\gamma \int_0^t [\boldsymbol{R}\boldsymbol{G}(\tau)]^\top \boldsymbol{x}_n d\tau\right\} d\boldsymbol{x}_n\right] \exp\left\{i\gamma \int_0^t [\boldsymbol{R}\boldsymbol{G}(\tau)]^\top \boldsymbol{r} d\tau\right\} \tag{4}$$

Then, the following two equations can be written:

$$s(t, \boldsymbol{G}(t)) = \hat{s}(t, \boldsymbol{R}^\top \boldsymbol{G}(t)) \exp\left\{-i\gamma \int_0^t \boldsymbol{G}(\tau) \boldsymbol{r} d\tau\right\} \tag{5}$$

Thus, the original signal can be recovered if an inverse of the rotation is applied onto the gradients and by applying a separate phase shift to the rotation corrected signal.

## 2.1 Fast Direct Rotation Calculation via Accelerometer and Magnetometer (MR-Compass)

In MRI, the permanent magnet creates a strong uniform magnetic field ($\boldsymbol{b}$) within a large region inside the bore. The Earth's gravitational field ($\boldsymbol{g}$) can be considered to be perpendicular to $\boldsymbol{b}$ everywhere within the bore. These fields in the North-East-Down (NED) coordinate frame are $\boldsymbol{a} = \begin{bmatrix} 0 & 0 & g \end{bmatrix}^\top$ and $\boldsymbol{b} = \begin{bmatrix} B_0 & 0 & 0 \end{bmatrix}^\top$. Then, the measurements from a stationary accelerometer and magnetometer will be as follows:

$$\boldsymbol{a}_m = \boldsymbol{R}\boldsymbol{g} \tag{6a}$$

$$\boldsymbol{b}_m = \boldsymbol{R}\boldsymbol{b} \tag{6b}$$

where $\boldsymbol{R} \in \mathbb{R}^{3\times 3}$ is the rotation matrix that represents the orientation of the sensor with respect to NED. Then, the orientation of the sensor can be calculated as follows [56]:

$$\boldsymbol{R}_{\text{NED}} = \left[ \begin{array}{c|c|c} \dfrac{(\boldsymbol{a}_m \times \boldsymbol{b}_m) \times \boldsymbol{a}_m}{g^2 B_0} & \dfrac{\boldsymbol{a}_m \times \boldsymbol{b}_m}{g B_0} & \dfrac{\boldsymbol{a}_m}{g} \end{array} \right] \tag{7}$$

Converting NED to the MRI device coordinate system (DCS) is a simple permutation:

$$\boldsymbol{R}_{\text{DCS}} = \boldsymbol{R}_{\text{NED}} \begin{bmatrix} 0 & 0 & -1 \\ 1 & 0 & 0 \\ 0 & -1 & 0 \end{bmatrix} \tag{8}$$

The calculated rotation matrix is the relative orientation of the sensor with respect to DCS. However, the initial attachment of the sensor onto the human body may not be aligned to DCS. This requires normalization with the initial rotation estimate as follows:

$$\boldsymbol{R}(t) = \left( \boldsymbol{R}_{\text{DCS}}^\top(0) \boldsymbol{R}_{\text{DCS}}(t) \right)^\top \tag{9}$$

$\boldsymbol{R}(t)$ can now be used within Eq. (5) to correct for the rotation prospectively or retrospectively. This normalization also acts as a self-calibration step that seamlessly solves the hand–eye coordination during the first pass of the method. Hence, the hand–eye coordination is solved at the system's sampling rate, and does not require additional scans or steps nor is patient specific. Once the rotation is calculated and applied, translation reduces to a phase difference which can be calculated via phase correlation from the MRI data[53, 57–60].

## 3 Methods

### 3.1 Sensor Calibration

For accurate estimations, sensor calibration is crucial. The calibration is done in two steps:

**i. Internal Misalignment (IM):** The individual sensor's internal axes may misalign due to manufacturing errors, resuling in non-orthogonal axes [61–65]. This necessitates recalibration using known references. As an example, the measurement model for a 3-axis accelerometer is as follows [61–66]:

$$\boldsymbol{a}_{im} = \boldsymbol{C}_a \boldsymbol{a}_m + \boldsymbol{c}_a + \boldsymbol{n} \tag{10}$$

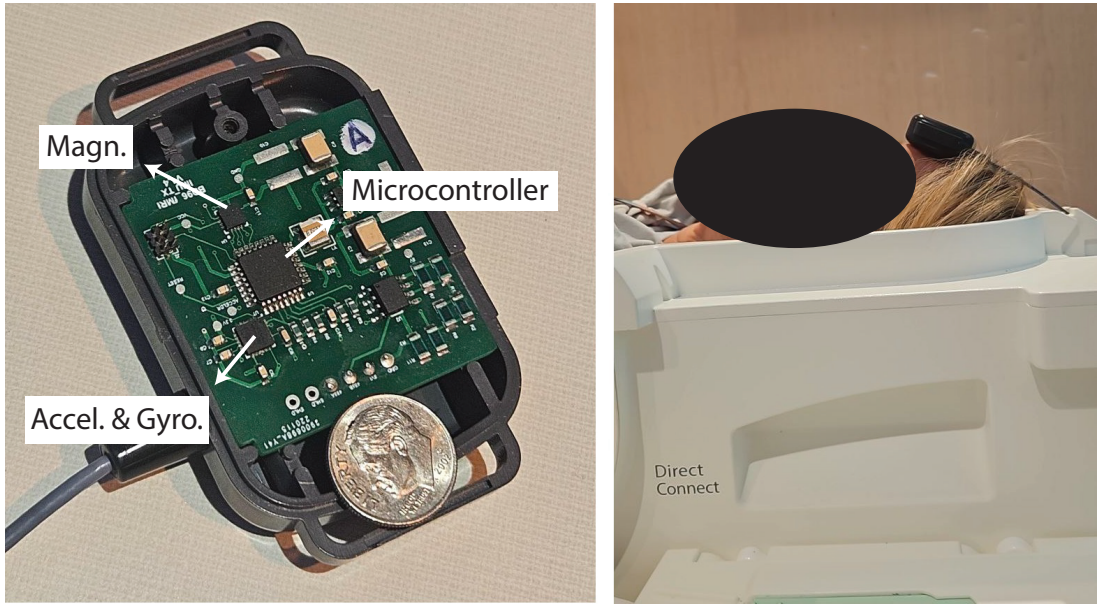

Figure 1: Sensor PCB with magnetometer (magn.), accelerometer (accel.) and gyroscope (gyro) with the microcontroller that aggregates the data and communicates with the digital signal processing box.

In Eq. (10), $\boldsymbol{a}_m$ is perturbed by a symmetric misalignment matrix (not a rotation matrix) $\boldsymbol{C}_a \in \mathbb{R}^{3\times 3}$, a bias vector $\boldsymbol{c}_a \in \mathbb{R}^{3\times 1}$, and Gaussian noise $\boldsymbol{n} \sim \mathcal{N}(\boldsymbol{0}, \sigma \boldsymbol{I})$. Then, the unknowns, $\boldsymbol{C}_a$ and $\boldsymbol{c}_a$, can be solved using the gravity as reference and a least squares approach. The same methodology is utilized to calibrate the magnetometer by taking the $B_0$ field as the reference.

**ii. Cross Misalignment (CS):** The orientation difference between a magnetometer and accelerometer (cross misalignment), can be resolved using the magnetic inclination angle ($\delta$) [66–68]. $\delta$ is defined as the angle between the horizontal plane and the magnetic field lines [66–68]:

$$\sin(\delta) = \frac{\boldsymbol{a}^\top \boldsymbol{R}_{cs} \boldsymbol{b}}{g B_0} \qquad (11)$$

The cross-sensor misalignment rotation matrix, $\boldsymbol{R}_{cs} \in \mathbb{R}^{3\times 3}$, can be solved using a minimization algorithm if $\delta$ is known. In the MRI bore, magnetic and gravitational fields are perpendicular everywhere around the isocenter, $\delta = 0°$. Then, the unknown, $\boldsymbol{R}_{cs}$, can be solved using a similar least squares approach. The Eqs. (10) and (11) indicate that both misalignments can be calculated with measurements from 12 distinct orientations.

These calibration steps are inherent to the sensor device itself and are not dependent on orientation of the device. Once calibrated, the sensor does not need recalibration unless a catastrophic event occurs.

**Algorithm 1** FFT-Based 2D slice-to-volume phase localization and 3D volume reconstruction.
---
    **Input:** New 2D slice $f$, Reference 3D volume $I$
    **Output:** Reconstructed 3D volume $I_m$
    **Initialize:**
    Generate refined z-direction:
        $z_{\text{int}} \leftarrow 10\times$ finer sampling of original z-grid
    Interpolate reference volume along z-direction only:
        $I_{\text{int}} \leftarrow \text{Interp}_z(I, z_{\text{int}}, \text{``spline''})$
    Compute 2D FFTs:
        $F_I \leftarrow \mathcal{F}_2(I)$
        $F_{I_{\text{int}}} \leftarrow \mathcal{F}_2(I_{\text{int}})$

1: **for** each new 2D slice f in a volume **do**
        $F_f \leftarrow \mathcal{F}_2(f)$
2:     **for** each slice index $k$ in $F_I$ **do**
3:         $C_k \leftarrow \Re\left\{\mathcal{F}_2^{-1}\left(F_I^{(k)} \odot F_f^*\right)\right\}$
4:     **end for**
5:     Determine coarse z-location:
        $k_{max} \leftarrow \arg\max_k (\max C)$
6:     Define refined search window:
        $\mathcal{Z} \leftarrow \{k \mid |k - k_{max}| \leq 5\}$
7:     **for** each slice index $k$ in $\mathcal{Z}$ **do**
8:         $C_k^{\text{int}} \leftarrow \Re\left\{\mathcal{F}_2^{-1}\left(F_{I_{\text{int}}}^{(k)} \odot F_f^*\right)\right\}$
9:     **end for**
10:     Determine refined z-location:
        $k_{int,max} \leftarrow \arg\max_k (\max C^{\text{int}})$
        $z_r \leftarrow z_{int}(k_{int,max})$
11:     Determine in-plane displacement:
        $(x^*, y^*) \leftarrow \arg\max_{(x,y)} \text{FIT}_2\left\{C_{z_r}^{\text{int}}(x, y)\right\}$
12:     Apply phase shift to slice:
        $f_{\text{aligned}} \leftarrow \text{PhaseShift}(f, x^*, y^*)$
13:     Insert aligned slice into moving volume:
        $I_m(:, :, z_r) \leftarrow \Re\{f_{\text{aligned}}\}$
14: **end for**
15: Interpolate aligned volume to original z-grid:
        $I_m \leftarrow \text{Interp}_z(I_m, z)$
---

## 3.2 Imaging Sequences

It is important to simultaneously characterize the error of MR-compass for completeness. Imaging sequences with inherent self-navigation capabilities provide an excellent opportunity for both error characterization and phase correlation. Therefore, we utilized the following two sequences:

**i. 3D radial**[33, 69, 70] is ideal for generating low-resolution images from subsets of 3D-radial k-space lines (spokes) to serve as reference scans. The rotations of the spokes are corrected via MR-compass, then, low-resolution 3D volumes are generated from sets of N-spokes (a shot). These low-resolution 3D volumes have the same orientation, however, different translations. The translation is then calculated in k-space via 3D phase correlation. Finally, the calculated translations are applied to the spokes as phase shifts and the final high-resolution volume is reconstructed. The 3D-radial sequence was used to acquire T1-weighted images (TR/TE = 6.86/3.05 ms, FA = $4^o/6^o$, FOV = 210x210x210 mm, 0.82 mm isotropic resolution, RBW = 260 Hz/px, 100 shots, 448 spoke/shot) over 6 minutes 20 seconds. Additionally, a mutual information-based iterative image registration (MI-Reg) algorithm is utilized to provide the motion correction baseline [33, 35, 69].

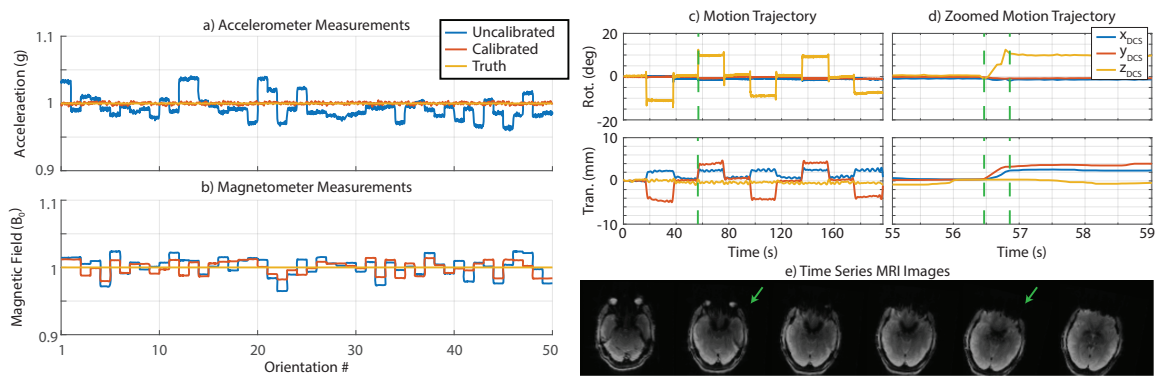

Figure 2: Norms of the accelerometer and magnetometer after calibration shows greater agreement with the gravity (a) and the field magnetic field (b) within the MRI bore compared to uncalibrated case. The improvement in the accelerometer is exceptionally high whereas magnetometer improvement is marginal. Example of motion onset capture for the fifth participant during a head-shake fMRI scan. The large motion, occurring in between the green lines in (c-d), is a very rapid event as the participant turns their head in approximately 400 ms. The green arrows in (e) correspond to the green lines in the zoomed figure. The MR-Compass method accurately reacts to the motion and measures the change.

**ii**. **2D EPI fMRI** sequence acquires 2D slices very fast and stacks them to generate volumes. An fMRI sequence was modified to accept motion parameters every slice. The sequence is then bombarded with the rotation parameters at 200 Hz, down from 2000 Hz through 10-fold averaging to reduce noise, in real-time. A total of 50 volumes were acquired with the fMRI sequence (TR = 4s, TE = 28ms, FA = $90^o$, FOV = 258x258x180 mm, 3 mm isotropic resolution, 60 slices per volume, 66.7 ms/slice) over 3 minutes 20 seconds.

### 3.3 Prospective Correction Pipeline

The IMU sensor shown in Fig. 1 (BIOPAC Systems Inc.) is connected to a digital signal processing box that aggregates the data and relays it to a PC in the MR computer room. This PC runs BIOPAC's ACQKNOWLEDGE software to acquire the data in real-time. The measurements are then sent to a high-performance processing server via TCP/IP.

The PMC pipeline runs on the server, applying Equations (7)-(9) in real-time, and communicates with the scanner via the Siemens FIRE framework prototype (Chow et al., 2019). It utilizes Siemens' prototype MoCo Framework for PMC. FIRE facilitates communication between the MRI system and external servers, managing the transfer of MRI-related data.

The pipeline functions in a fully integrated feedback loop. Communication with the MRI system and sensor is established at startup. Once the sequence starts, fMRI images are streamed to the pipeline, which immediately begins providing orientation parameters to the sequence upon receiving

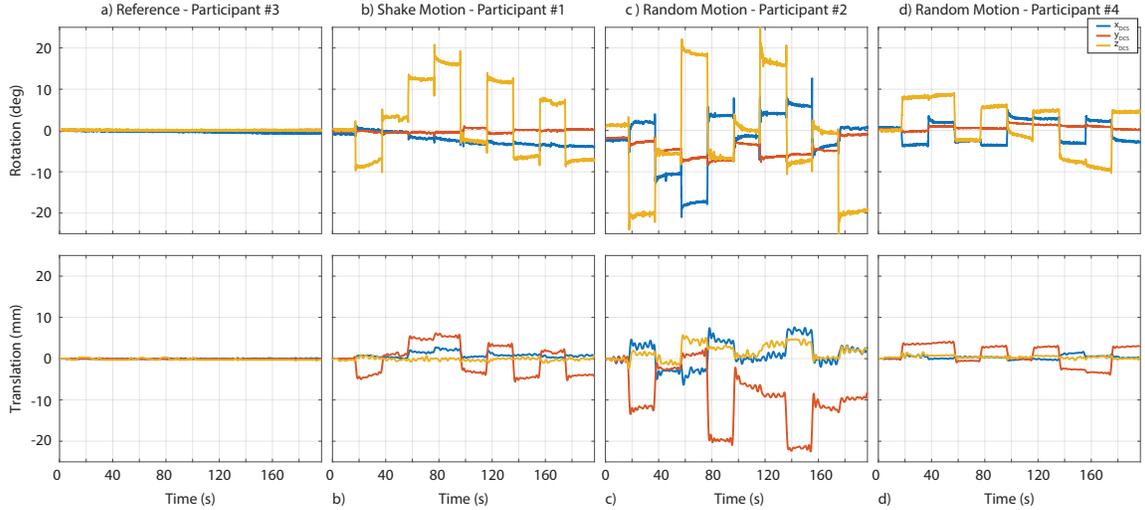

Figure 3: Example motion trajectories for different participants and motion types for the fMRI experiments. Rotations are tracked by the MR-compass at 200 Hz whereas the translations are calculated using slice-to-volume phase correlation as described in Algorithm 1. The participants were instructed to move approximately every 5 volumes and the motion events were captured accurately with the MR-compass. The motion traces include a mixture of high magnitude repositioning and drifts, covering a large range possible motions in an MRI scan.

the first data packet.

The resulting fMRI slices have corrected orientations but still require translation correction. Phase correlation is a well-established method for subpixel shift estimation [71, 72]. These algorithms were extended in a custom toolbox to perform slice-to-volume phase correlation in k-space (Algorithm 1).

The algorithm uses a two-tier approach: coarse localization along the z-axis followed by fine localization near the coarse estimate. Standard 2D phase-correlation methods [71, 72] then estimate subpixel shifts along x and y. The input slice is phase-shifted along the x- and y-axis and the z-position is updated. After all slices are processed, a new volume is resampled onto the original z-grid.

The modified fMRI sequence accepts motion parameters for each individual slice. Thus, the PMC pipeline is also compatible for simultaneous multi-slice acquisitions.

### 3.4 Volunteer Experiments

Five volunteers were imaged at 3T (MAGNETOM Prisma; Siemens Healthineers, Forchheim, Germany) using a 64-channel head/neck coil with IMU attached to the forehead. For performance analysis, three scans were acquired per volunteer per sequence:

**i. Reference:** Volunteers held as still as possible, with correction active,

**ii. Head-Shake Motion:** Volunteers repeatedly repositioned their heads left-to-right in-plane,

**iii. Random Motion:** Volunteers repeatedly moved their heads randomly in any direction (in- or through-plane).

During the motion experiments, movement cues were delivered via intercom every 30 and 20

seconds for the 3D-radial and fMRI sequence, respectively. The restrictive padding around the volunteers was removed to enhance mobility. Five volunteers were scanned for the 3D-radial sequence and four for the fMRI sequence, totaling 27 cases.

For the 3D radial sequence, SSIM, normalized variance of the Laplacian (NVoL), and normalized mutual information (NMI) were used. NVoL values were normalized by the reference VoL. For fMRI, SSIM was used. All metrics were computed on brain-only images obtained via skull stripping [73–75].

### 3.5 Characterization of Sensor Error

Sensor readings are corrupted by noise, hence the estimated rotations will have errors. While a detailed derivation of the theoretical sensor error bounds are provided in Section 5.1, an experimental measurement of this error is crucial. The self-navigation capabilities of the chosen sequences are utilized through the well-established MI-Reg methods[35, 69, 76–79] to provide a validation benchmark.

For the 3D-radial sequence, low-resolution volumes from groups of 448 spokes (1 shot) of MR-Compass data are registered to the reference volume using the state-of-the-art MI-Reg[15, 33, 35, 69, 76, 77, 79] method to estimate residual 6-DOF motion parameters. Similarly, for the fMRI sequence, MR-Compass-corrected time-series fMRI images are registered individually to the reference volume with the MI-Reg method. The errors are then calculated as follows:

$$\epsilon_{\boldsymbol{R}} = \cos^{-1}\left(\frac{\mathrm{Tr}(\boldsymbol{R}_{\mathrm{res}}) - 1}{2}\right) \tag{12a}$$

$$\epsilon_{\boldsymbol{r}} = \|\boldsymbol{r}_{\mathrm{res}}\|_2 \tag{12b}$$

where $\epsilon_{\boldsymbol{R}}$ and $\epsilon_{\boldsymbol{r}}$ denote the norms of the residual axis angle and translation, respectively. Here, $\boldsymbol{R}_{\mathrm{res}}$ and $\boldsymbol{r}_{\mathrm{res}}$ are the residual difference in orientation and translation.

## 4 Results

### 4.1 Sensor Calibration

To reduce noise, measurements from 200 random orientations were averaged for 1.5 s per orientation at a 2 kHz sampling rate, yielding an SNR improvement of $10 \log_{10}(\sqrt{3000}) = 17.4$ dB. An additional 50 measurements were collected for validation. Figure 2.a-b shows the calibration performance on the validation data. The accelerometer calibration performs exceptionally well, with the calibrated measurements (red) nearly matching the ground truth (orange). The magnetometer calibration reduces overall variance but is less accurate. The MSE improves by ∼200× and ∼2.3× for the accelerometer and magnetometer, respectively. The expected rotation estimation error from the residual magnetometer misalignment is minimal, as the calibrated magnetometer has a standard deviation of 0.03 T, only 1% of $B_0$. The cross-sensor misalignment angle is estimated as $\boldsymbol{\theta}_{cs} = \begin{bmatrix} -0.003 & -0.146 & 0.1277 \end{bmatrix}^o$.

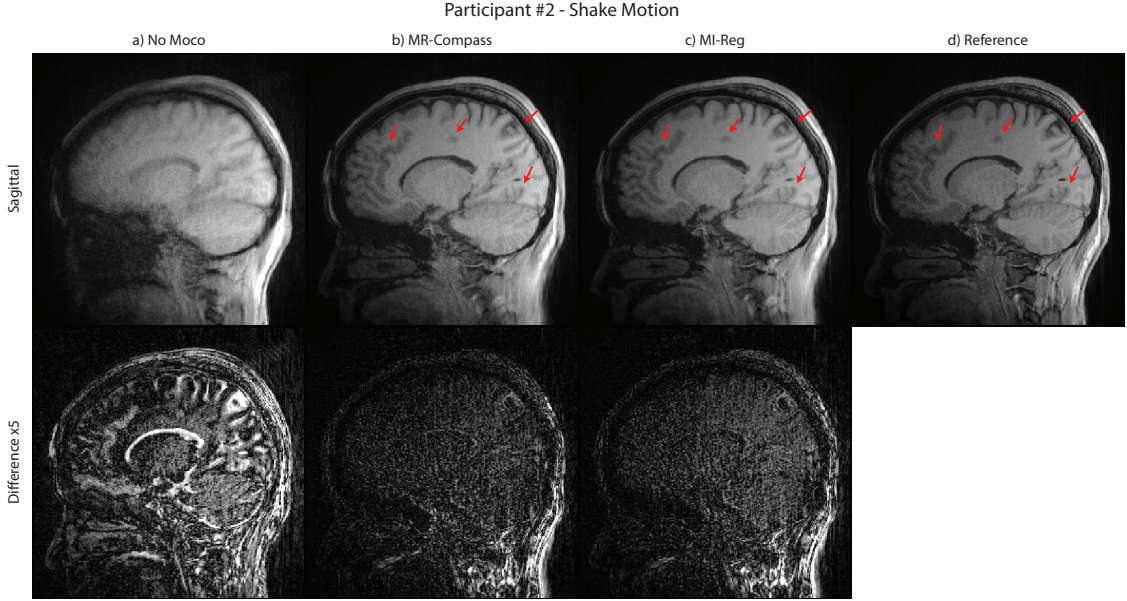

Figure 4: The results of the 3D-radial sequence reconstructions for the shake motion of participant 2. The proposed MR-compass and phase correlation shows excellent image reconstruction performance as can be seen from the selected areas shown with the red arrows. The differences of MR-Compass and MI-Reg images are practically the same as can be seen from their difference images where the amplified error images do not exhibit any structure.

### 4.2 Detection of Onset of Motion

Motion onset is illustrated in Fig. 2.c-e for the fourth volunteer during the head-shake fMRI scan. A motion event, marked by green vertical lines, begins at approximately 56.465 s and ends at 56.86 s, for a total duration of 395 ms. Motion-corrupted images at the start and end of the event are indicated with arrows. The event spans about six slices (∼400 ms), demonstrating that the high temporal resolution MR-Compass captures motion accurately and immediately.

The total end-to-end delay of the PMC pipeline is $50 \pm 3$ ms, measured using the wall clocks of the acquisition PC and the MR PC, which is slightly less than the acquisition time of a single slice (66.7 ms). The total processing time of the MR-Compass method for one measurement is $80 \pm 4$ $\mu$s, well below the 5 ms (200 Hz) sample rate. These timings were achieved on a server with a 32-core AMD EPYC 7542 processor.

For RMC, synchronization is achieved using the wall clock of the sensor measurement PC and the acquisition date–time information embedded in the raw MRI data.

### 4.3 Motion Trajectories

The fMRI motion trajectories are shown in Fig. 3. Orientations are captured by the MR-Compass and the translations are estimated via slice-to-volume phase-correlation algorithm. In Fig. 3a, although the participant was instructed to remain still, gradual drift occurred and was accurately captured by the MR-Compass.

During the head-shake and random-motion scans, participants were instructed to move as much as possible within the head coil, resulting in large rotations (up to $\pm 20°$) and translations (up to $\pm 20$ mm). In addition to these major motions, substantial drift is also observed, as shown in Figs. 3.b (x-axis rotation), 3.c (y-axis translation), and 3.d (z-axis rotation). Similar motion patterns were observed during the 3D-radial sequence. These trajectories demonstrate that the proposed system captures motion with high fidelity.

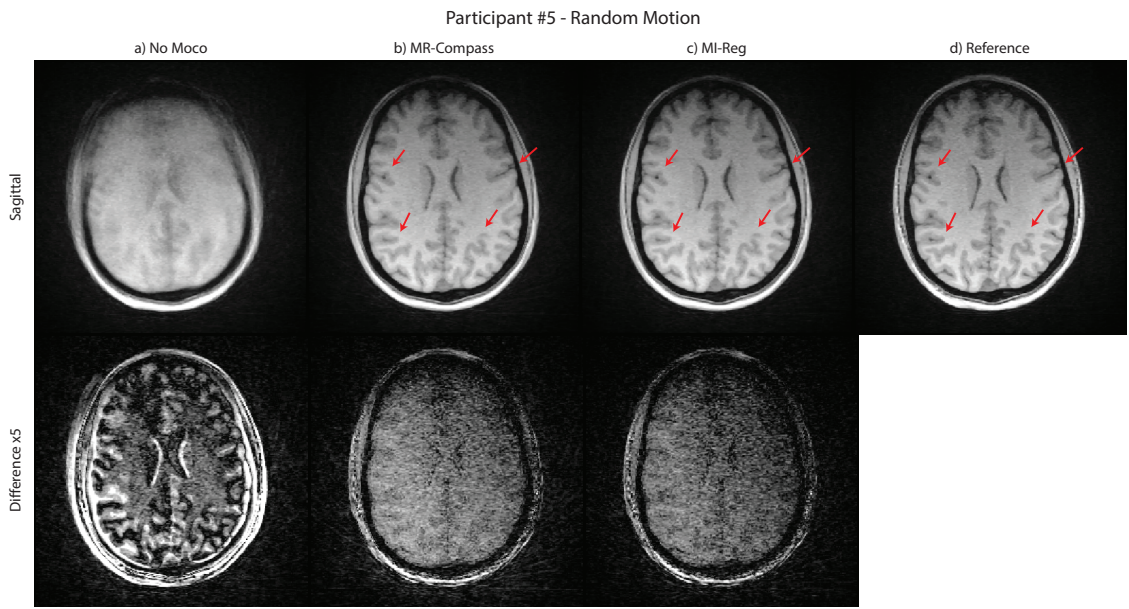

Figure 5: The results of the 3D-radial sequence reconstructions for the random motion of participant 5. The proposed MR-compass and phase correlation shows excellent image reconstruction performance as can be seen from the selected areas shown with the red arrows. The differences with respect to the reference show very similar noise-like behavior for MR-compass and MI-Reg, whereas the no correction image has considerably more error.

## 4.4 Retrospective Correction Performance

Figures 4 and 5 shows representative examples of sagittal and axial slices through uncorrected, retrospectively corrected with MR-compass and MI-Reg, and reference 3D-radial images with head shake and random motion, respectively. Substantially reduced blurring can be seen with both MR-Compass motion correction. The difference images show both MR-compass and MI-Reg result in practically the same.

Figure 6 shows the mean SSIM, NVoL and NMI values among the participants for different motion paradigms. For all the cases, both the MR-Compass and MI-Reg methods show very good agreement and the no correction results are significantly worse, reflecting the quality difference in the images in Figures 4 and 5. Furthermore, the MR-Compass and MI-Reg result in very similar reconstructions. The percentage difference between the image metrics of MR-Compass and MI-Reg are 0.4%, 1.5% and 1.1% for the head-shake motion, and 0.2%, 2% and 0.8% for the random motion paradigms, respectively. This is also reflected in the difference images in Figures 4 and 5.

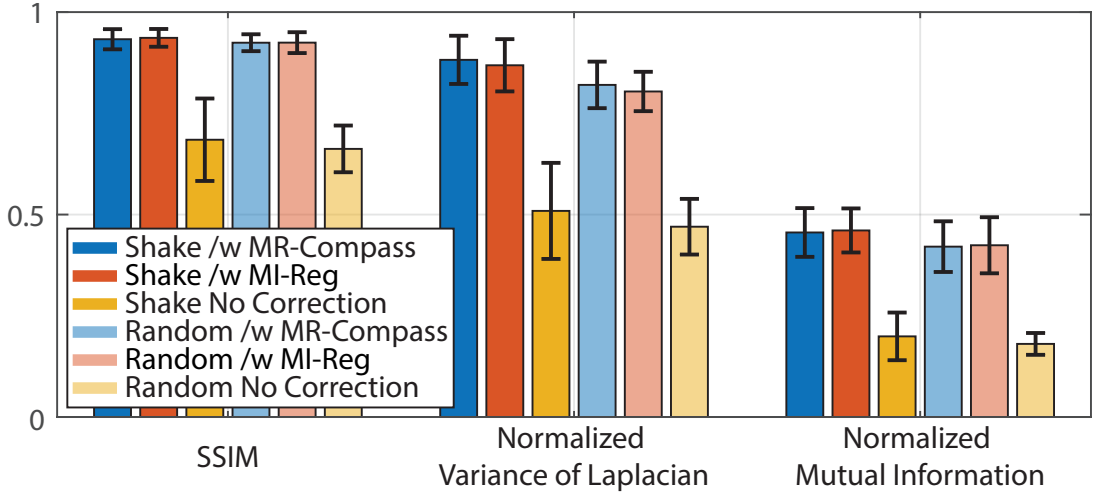

Figure 6: Mean and standard deviation of image statistics across participants for different motion paradigms show that the proposed MR-Compass has similar performance to state-of-the-art MI-Reg.

### 4.5 Prospective Correction Performance

Figure 7.a shows the average SSIM values per volume for the third participant using the MR-Compass method under no-motion, shake, and random-motion conditions. The no-motion scan confirms that the method does not degrade image quality; it actually captures and corrects the participant's slow drift (Fig. 3.a). During both motion paradigms, the method consistently maintains image alignment with the reference.

Figure 7.b presents example slices, showing MR-Compass difference images have significantly lower error. The residual errors present in the MR-Compass difference images result from contrast changes related to the receive-coil profiles as the participant moves a phenomenon known as "intensity modulation" [80–82].

Figure 8.a shows that the average SSIM among the participants is very close to the reference for each volume in the time series, above 0.94 with minimal variations. The performance of the MR-Compass is slightly lower for random motion compared to shake motion. Figure 8.b shows the mean and standard deviation of SSIM averaged over all volumes for each participant and motion type separately. The SSIM values are very close to the reference and the low standard deviations also reveal that the performance does not change over time.

The slice-to-volume phase correlation algorithm calculates slice translation in the x-, y-, and z-directions in $15 \pm 1.8$ ms per slice, substantially faster than the slice acquisition time. The algorithm is implemented in MATLAB as a post-processing step on a PC with an 8 core AMD Ryzen 7 Pro 7735U processor.

### 4.6 Motion Parameter Estimation Error

Figure 9 presents these motion parameter estimation errors of MR-Compass. Both the rotational and translational trajectory error is overall slightly higher in 3D-radial sequence compared to fMRI. Furthermore, the random head shake motion has higher rotational and translational error across all the experiments.

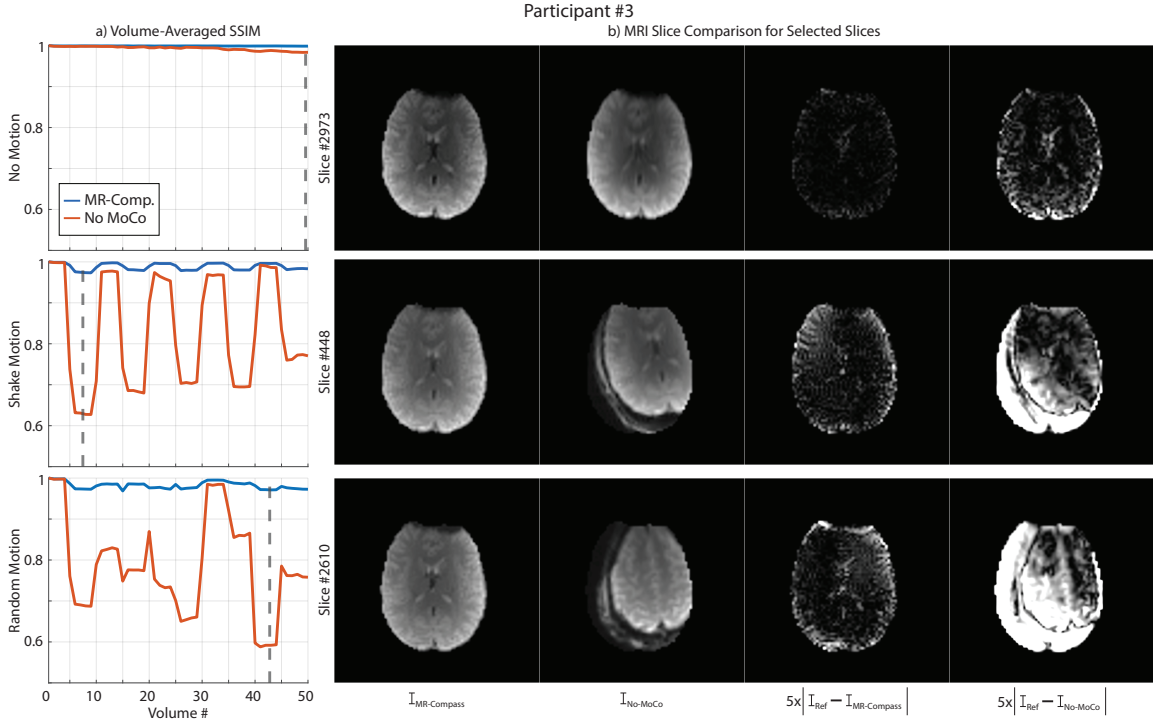

Figure 7: The proposed MR-compass and slice-to-volume phase correlation for fMRI sequence shows excellent image alignment performance as can be seen from time series SSIM plots and fMRI images of selected slices on these plots (shown with gray dashed lines). The difference images demonstrate the effectiveness of the proposed MR-Compass method. When compared to the reference image, the remainder is noise-like without any clear structure. The no-motion images are obtained through corrupting the images with the rotation parameters obtained by MR-compass.

For the 3D-radial case, the mean error of the trajectory among all participants for the head-shake motion is $0.81^o \pm 0.43^o$ and $0.55 \pm 0.45$ pixel (1 mm resolution), and random motion is $0.96^o \pm 0.45^o$ and $0.61 \pm 0.46$ pixel (1 mm resolution).

For the fMRI case, the mean error of the trajectory among all participants for the head-shake motion is $0.52^o \pm 0.38^o$ and $0.24 \pm 0.25$ pixel (3 mm resolution), and random motion is $0.69^o \pm 0.38^o$ and $0.32 \pm 0.23$ pixel (3 mm resolution).

## 5 Discussion

In this work, we have proposed and experimentally demonstrated techniques that enable an IMU-based image alignment for both RMC and PMC for two fundamentally different sequences. The proposed MR-Compass method provides excellent robustness against a wide variety of motions, that include major repositions and also drifts. The method delivers results at an unprecedented 200 Hz sampling rate, down from 2000 Hz through 10-fold averaging to reduce noise, and accuracy that provides very close image quality to the reference. Importantly, the proposed method performs comparatively to the state-of-the-art MI-Reg.

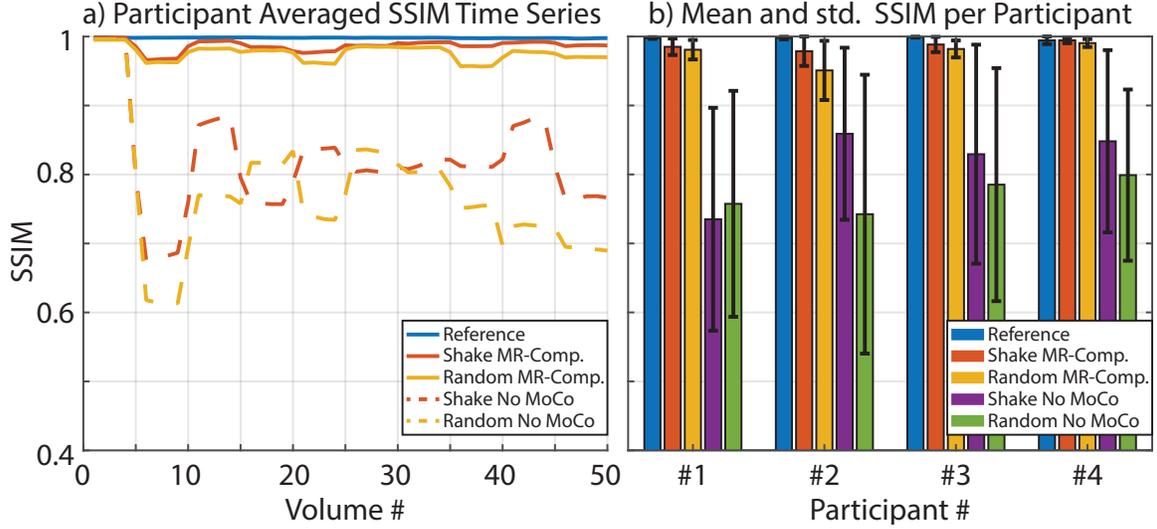

Figure 8: Image quality statistics of the fMRI experiments for (a) average SSIM with respect to participants for each time point for each motion type, (b) average SSIM with respect to time for each participant for each motion type.

## 5.1 Trajectory Estimation Error

The experimental trajectory errors are presented in Section 4.6. A theoretical orientation error can be formulated using infinitesimal rotations and exponential maps[83]:

$$\boldsymbol{R}_n = \boldsymbol{R}\exp([\boldsymbol{\epsilon}]^\wedge) \qquad (13)$$

where $\boldsymbol{R}$ is the noise-free rotation matrix that represents the orientation of the sensor, $\boldsymbol{\epsilon} = \begin{bmatrix} \alpha & \beta & \gamma \end{bmatrix}^\top$ is a small angle perturbation and $[\cdot]^\wedge$ denotes transformation of the underlying vector to a skew-symmetric matrix. First order Taylor series expansion results in:

$$\boldsymbol{R}^\top \boldsymbol{R}_n = \boldsymbol{I} + [\boldsymbol{\epsilon}]^\wedge = \begin{bmatrix} 1 & -\gamma & \beta \\ \gamma & 1 & -\alpha \\ -\beta & \alpha & 1 \end{bmatrix} \qquad (14)$$

$\boldsymbol{R}^\top \boldsymbol{R}_n$ can be extended using the noiseless rotation matrix in Eq. (7) and its noisy versions:

$$\boldsymbol{R}^\top \boldsymbol{R}_n = \begin{bmatrix} \boldsymbol{r}_1^\top \\ \boldsymbol{r}_2^\top \\ \boldsymbol{r}_3^\top \end{bmatrix} \begin{bmatrix} \dfrac{(\boldsymbol{a}_n \times \boldsymbol{b}_n) \times \boldsymbol{a}_n}{\|(\boldsymbol{a}_n \times \boldsymbol{b}_n) \times \boldsymbol{a}_n\|} & \bigg| & \dfrac{(\boldsymbol{a}_n \times \boldsymbol{b}_n)}{\|\boldsymbol{a}_n \times \boldsymbol{b}_n\|} & \bigg| & \dfrac{\boldsymbol{a}_n}{\|\boldsymbol{a}_n\|} \end{bmatrix} \qquad (15)$$

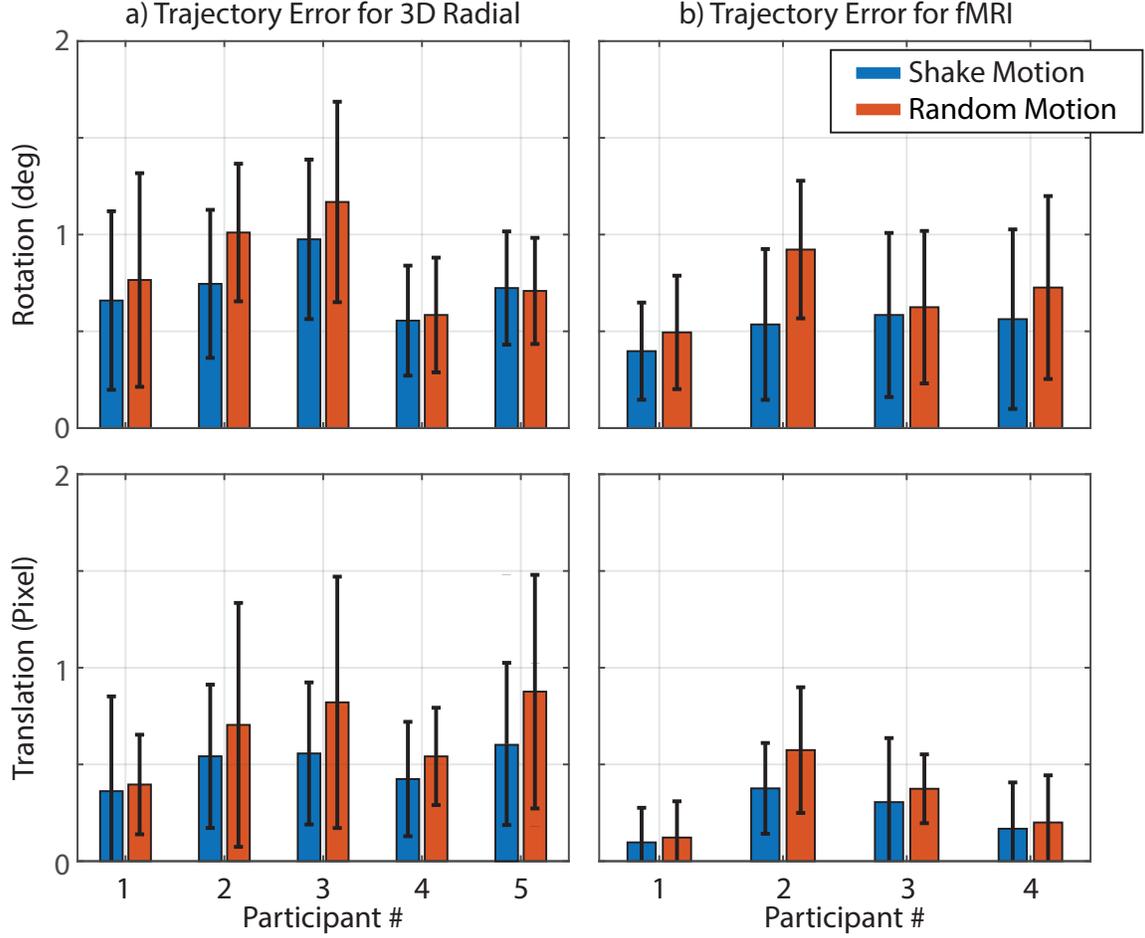

Figure 9: The trajectory errors for (a) the 3D-radial and (b) fMRI experiments show that the proposed MR-Compass method captures the motion with very good accuracy.

The small angle perturbations can be calculated as follows:

$$\boldsymbol{\epsilon} = \left[ \begin{array}{c|c|c} -\boldsymbol{r}_2^\top \dfrac{\boldsymbol{a}_n}{\|\boldsymbol{a}_n\|} & \boldsymbol{r}_1^\top \dfrac{\boldsymbol{a}_n}{\|\boldsymbol{a}_n\|} & -\boldsymbol{r}_1^\top \dfrac{\boldsymbol{a}_n \times \boldsymbol{b}_n}{\|\boldsymbol{a}_n \times \boldsymbol{b}_n\|} \end{array} \right]^\top \tag{16}$$

Eq. (16) reveals the dependence of the angular noise to $\boldsymbol{R}$. Thus, the sensor noise is no longer independent if $\boldsymbol{R} \neq \boldsymbol{I}$. Furthermore, the denominator becomes the square root of generalized $\chi^2$ with $k = 3$ degrees of freedom, which does not have a closed form solution. For this reason, the sensor readings are assumed to be perturbed by small values as follows[84]:

$$\begin{aligned} \boldsymbol{a}_n &= g\boldsymbol{R}\boldsymbol{e}_3 + \Delta\boldsymbol{a} \\ \boldsymbol{b}_n &= B_0\boldsymbol{R}\boldsymbol{e}_1 + \Delta\boldsymbol{b} \end{aligned} \tag{17}$$

where $\Delta\boldsymbol{a}, \Delta\boldsymbol{b} \in \mathbb{R}^{3\times 1}$ are the perturbations and $\boldsymbol{I} = \left[ \begin{array}{c|c|c} \boldsymbol{e}_1 & \boldsymbol{e}_2 & \boldsymbol{e}_3 \end{array} \right]$ are the unit vectors. Combining

Eqs. (17) and (16) yields:

$$\alpha = -\boldsymbol{e}_2^\top \boldsymbol{R}^\top \frac{\Delta \boldsymbol{a}}{\|\boldsymbol{a}_n\|} \tag{18a}$$

$$\beta = \boldsymbol{e}_1^\top \boldsymbol{R}^\top \frac{\Delta \boldsymbol{a}}{\|\boldsymbol{a}_n\|} \tag{18b}$$

$$\gamma = \frac{g\boldsymbol{e}_2^\top \boldsymbol{R}^\top \Delta \boldsymbol{b} - \boldsymbol{e}_1^\top \boldsymbol{R}^\top [\Delta \boldsymbol{a}]^\wedge \Delta \boldsymbol{b}}{\|\boldsymbol{a}_n \times \boldsymbol{b}_n\|} \tag{18c}$$

One can show, numerically, that the $\alpha, \beta, \gamma$ angles obtained from either the upper or lower triangle of the error matrix in Eq. (14) will be very close to each other for small $\Delta \boldsymbol{a}, \Delta \boldsymbol{b}$. This formulation also provides an opportunity to embed the sensor calibration errors as well, since calibration matrices and bias vectors from Equations (10) and (11) will add to $\Delta \boldsymbol{a}, \Delta \boldsymbol{b}$. Furthermore, it also reveals that the error on $\gamma$ will be less compared to the error on $\alpha$ and $\beta$, which can be observed in Fig. 3, where $\gamma$ corresponds to the y-axis.

The standard deviations of the accelerometer and magnetometer are measured as 0.05 $m/s^2$ and 0.0012 T per channel at 200 Hz, respectively. Assuming $\boldsymbol{R} = \boldsymbol{I}$, Eq. (18a)-(18c) results in an angular error norm of $0.41^o$. Comparing this to the measured angular errors shows exceptional agreement. The remaining errors can be associated to changing orientation $\boldsymbol{R}(t)$ and other additional effects, such as sensor nonlinearities, and calibration errors.

Equation (8) assumes no kinematic acceleration of the rigid body. Our study found that most kinematic acceleration occurs briefly at the start and end of the motion event. Additionally, our measurements show that the human head's kinematic acceleration is typically small (Supplementary Material 1). Therefore, the error from kinematic acceleration is negligible. This is further supported by the closeness of the MR-Compass and MI-Reg methods for the 3D-radial sequence.

The effects of kinematic acceleration and noise can be reduced with low-pass filtering. However, in this work, filtering was avoided to prioritize system responsiveness over noise reduction, as filters could introduce group delays and pass-band ripples that lead to estimation errors.

Both theoretical analysis and experimental results suggest that MR-Compass performance is mainly constrained by sensor quality. Better sensors with enhanced noise and linearity features would significantly improve accuracy without altering the methodology.

## 5.2 Temporal Resolution of Motion Parameter Estimation

For PMC, the interaction of the motion estimator system and the sequence is upper-bounded by the ability to apply motion parameters to the sequence. If time-varying $R(t)$ and $\boldsymbol{r}(t)$ could be estimated continuously, they could correct the sequence. However, MR controllers (MRCs) do not support continuous updates. An MRC defines a sequence as a series of discrete event blocks (EBs) and EBs can only be updated before entering the digital signal processor of the MR system [25].

Traditionally, MR sequences are loop-based with one TR per EB, applying PMC once per TR. Modern MRCs allow a TR to be divided into significantly shorter tasks (e.g., RF, gradients, ADC, delays) as real-time event blocks (RTEBs) [25], enabling real-time sequence construction [25, 85]. This facilitates motion updates per RTEB, enhancing adherence to Bloch equations and improving image quality [25, 54]. Herbst et al. [54] noted that high-temporal-resolution tracking systems would greatly benefit sequences with shorter RTEBs by more accurately satisfying the Bloch equations.

The 200 Hz MR-Compass allows precise motion parameter application at the RTEB level, thus could benefit sequences with shorter TRs or other modified sequences that facilitate higher-rate motion updates.

## 5.3 Synchronization with the MRI Sequence

PMC requires precise synchronization and minimal delay to provide the most up-to-date motion parameters to the sequence, otherwise, the motion parameters will be out of date and will result in motion-corrupted images.

The synchronization with the MRI system is obtained by initiating the sensor acquisition through the data packets received from the MRI system. Although this method is sufficient for the tested fMRI sequence as shown in Section 4.5, an improvement could be made by obtaining a precise digital trigger from the MRI system. Such triggers are widely used to synchronize external sensors such as those utilized for fMRI to synchronize the tasks to the sequence.

Another consideration is the end-to-end communication delay between the sensor and the MRI system. In this work, the sensor first connects to a digital signal processing box, then to a measurement PC, followed by a high-performance processing server, and finally to the MRI system.

An improvement would be to rewire the sensor to eliminate the signal processing box. Furthermore, it may also be possible to run the MR-Compass method on an embedded chip on the sensor board, since the required operations are minimal. In that case, the measurement PC could also be removed, allowing the sensor to connect directly to the MRI system. These solutions would require a complete redesign of the sensor assembly, which is an important future work.

## 5.4 Attachment of the IMU onto the Head

Reviews of PMC [10, 86] have discussed various methods for attaching tracking targets to a subject. Common strategies include a mouthguard with dental molds [85], a headband [21, 87], a marker on glasses frames [88, 89], or a marker attached with medical-grade tape [12, 90]. Studies comparing these mechanisms conclude that each method can provide good results in specific imaging scenarios[10].

Marker-free approaches have also been explored, including MRI-safe ink [91] and tracking intrinsic skin features [26]. The Tracoline system tracks facial skin directly but is affected by nonrigid motions such as blinking, frowning, and eyebrow movement [18, 19].

The IMU was attached onto the center of the forehead, hence any motion that includes eyebrows could effect the measurements, similar to that of observed in optical systems that track the face or markers [18, 19]. However, the expected error from this effect is minimal, since the additional acceleration is proportional to the applied force and the forehead muscles are not expected to exert large forces.

Another concern is slippage of the sensor from its initial position. In this work, IMU was securely attached to the skin using medical-grade double-sided adhesive tape. In this configuration, both RMC and PMC results indicate that no slippage occurred even after major motion events.

## 5.5 Hardware Safety Concerns

Any electronic device inside the MR bore can adversely affect both patient safety, due to applied gradients and RF fields, and image quality, due to potential perturbations of the $B_0$ field. Boston

Children's Hospital requires a strict MRI safety testing. The sensor did not display artifacts on the test images and no additional heating was found (Supplementary Material 2).

The sensor uses twisted-pair copper cables for power and data. Any loops can act as antennas, allowing RF and gradient fields to induce currents that may corrupt the signal, cause heating or damage to the sensor. One solution is to use fiber-optic cables. However, they are stiff and may restrict patient motion. Large movements could also cause the cables to exert additional force on the sensor, leading to erroneous measurements. Another option is wireless communication [47], but this would require an onboard battery and may introduce additional safety concerns. Therefore, copper cables remain the most practical solution, provided that loops are avoided.

## 5.6 Differences between Earth Navigation and MR Navigation

### 5.6.1 Orientation

In the traditional Earth navigation problems, the orientation is estimated via integrating the angular velocity obtained from the gyroscope. The orientation is calculated recursively as follows:

$$\boldsymbol{R}(t + \Delta t) = \boldsymbol{R}(t)\exp([\boldsymbol{\omega}(t)]^\wedge \Delta t) \tag{19}$$

Gyroscope measurements are inherently noisy, and integration therefore leads to drift due to angular random-walk. In Earth navigation, this drift has to be periodically corrected; however, the compass method in Eq. (7) is generally not preferred. This is because the Earth's magnetic field is very weak, varies in magnitude and direction with location, and is easily distorted by nearby metallic objects.

In contrast, the magnetic field in MRI is extremely strong and can dominate any local distortions. Moreover, the sensor assembly contains only minimal electronics to meet safety requirements. Another key advantage is that, near the isocenter, the MRI magnetic field is perpendicular to gravity everywhere around a large volume, so the inclination angle does not vary with position, in contrast to Earth navigation. Consequently, the compass algorithm is well suited for rotation estimation in MRI, as it avoids angular random-walk.

### 5.6.2 Translation

Translation can be calculated directly with the IMU as well. The traditional approach is double-integration of the accelerometer measurements. However, the measured acceleration is corrupted by the orientation dependent gravitational acceleration and noise:

$$\boldsymbol{a}_m(t) = \boldsymbol{a}_r(t) + \boldsymbol{R}(t)\boldsymbol{g} + \boldsymbol{n} \tag{20}$$

where $\boldsymbol{a}_r(t)$ is the kinematic acceleration, $\boldsymbol{R}(t)$ is the orientation and $\boldsymbol{n}$ is a Gaussian noise process.

If $\boldsymbol{R}(t)$ is known, the effect of gravity can be subtracted, leaving only kinematic acceleration and noise. Double integration then produces a position estimate corrupted by random-walk noise. In Earth navigation problems, a high-rate ($\geq 100$ Hz) IMU is fused with a low-rate ($\sim 1$ Hz) GPS signal for periodic position fixes, achieving centimeter-level accuracy [40, 92, 93].

Errors in estimating $\boldsymbol{R}(t)$ introduce additional residual gravitational acceleration and degrade position estimates further. Without an absolute reference, an IMU quickly drifts from the true position. In MRI, no GPS-like reference exists, so translation estimates rely entirely on dead reckoning.

Consequently, accurate translation estimation using IMUs in MRI is currently infeasible given the millimeter-level precision required.

In this work, direct orientation estimation reduces translation to a phase difference that is then estimated from the MRI data. Phase correlation requires reference data, such as a low-resolution image for 3D radial sequences or an initial reference volume for fMRI. Consequently, this approach can be extended to any k-space or image data that can serve as a reference for phase-correlation–based translation estimation.

Such signals (e.g., k-space lines or low-resolution images) can therefore be viewed as GPS-like references within MRI, providing low–temporal-resolution but accurate positioning. A high-performance, low-noise IMU can then fill in between each GPS-like signal using a kinematic state-space model to estimate high–temporal-resolution motion parameters. A key direction for future work is the development of such GPS-like reference signals within MRI and their fusion into the sensor pipeline to enable reliable translation estimation from the IMU.

# 6 Conclusion

In this work, we propose an IMU-based motion estimation method for both retrospective and prospective motion compensation in MRI head scans. A low-cost sensor assembly comprising an accelerometer, gyroscope, and magnetometer directly estimates head orientation with high accuracy, avoiding random-walk drift. Remaining translations are recovered via fast phase correlation in k-space. The method operates without visual line-of-sight, unlike camera-based systems, at much higher rates than navigator-based methods, and without patient-specific calibration such as pilot tone. Its fast self-calibration avoids lengthy procedures and clinical workflow bottlenecks. The method was validated on a T1-weighted 3D radial structural sequence and an fMRI sequence, achieving image quality comparable to MI-Reg for the 3D radial data and accurate alignment for the fMRI time series. In vivo results demonstrate robustness to a wide range of motions, including rapid repositioning and continuous drift.

# Supplementary Material 1 – Norm of the Magnetometer During Motion and Its Effect on the Motion Parameter Estimation

The sensor readings are assumed to be perturbed by small values were modeled as follows:

$$a_n = gRe_3 + \Delta a$$
$$b_n = B_0 Re_1 + \Delta b$$

Then, the residual angular errors can be formulated as follows:

$$\alpha = -\frac{e_2^T R^T \Delta a}{\|a_n\|}$$
$$\beta = \frac{e_1^T R^T \Delta a}{\|a_n\|}$$
$$\gamma = \frac{g e_2^T R^T \Delta b - e_1^T R^T [\Delta a]^\wedge \Delta b}{\|a_n \times b_n\|}$$

The figure below shows the norm of the accelerometer measurements and estimated rotations from the 5th participant's random motion experiment for the 3D-radial sequence. The accelerometer's norm has a mean difference of $0.01g$ relative to gravity, indicating a 1% residual error.

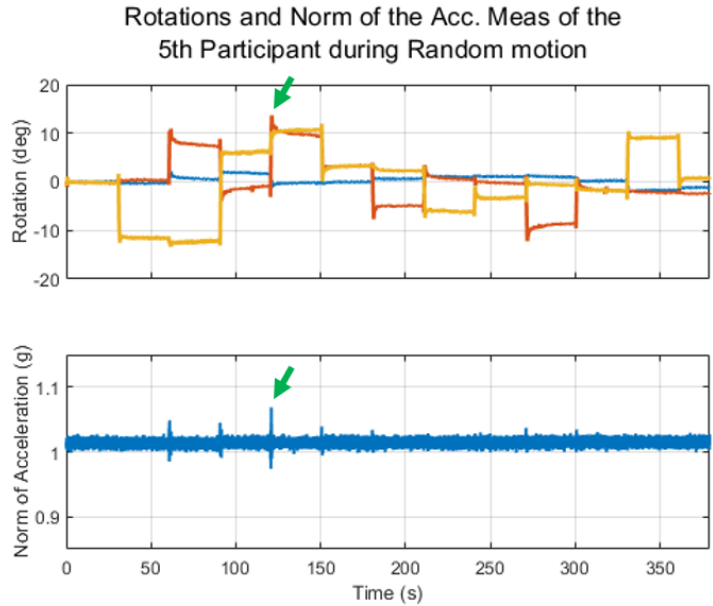

Figure 1. The norm of the accelerometer measurements and the estimated rotations from the 5[th] participant's head-shake motion experiment for the 3D-radial sequence indicate that the human head motion involves significant angular change but does not display large accelerations, only reaching about 0.06g at the worst case. Green arrows show the effect of kinematic acceleration during the motion event.

One can then calculate an average $\Delta a$, assuming that all axes of the accelerometer have identical perturbation:

$$\|\Delta a\| = \sqrt{\Delta a_x^2 + \Delta a_y^2 + \Delta a_z^2} = \Delta a_x \sqrt{3}$$

Plugging in $0.01g$ results in:

$$\Delta a_x \sqrt{3} = 0.01$$

$$\Delta a_x = \Delta a_y = \Delta a_z = 0.005g$$

Then adding this value to the already measured Gaussian noise std of $0.005g$ per axis, using the measured Gaussian noise std of $0.0012T$ of the magnetometer and plugging in these values into the angle error formula above results:

$$[\alpha \quad \beta \quad \gamma] = [-0.57 \quad 0.57 \quad 0.02]^o$$

$$\|[\alpha \quad \beta \quad \gamma]\| = 0.8^o$$

This closely aligns with the measured angular errors presented in the results. Moreover, Figure 1 shows that although the human head motion involves substantial angular change, it does not exhibit large accelerations. During the motion events, momentary kinematic accelerations occur due to the motion of the head (marked with green arrows on Figure 1). These can reach as much as $0.06g$. However, these momentary peaks settle down in around 5-6 ms in a motion event that spans 300-400 ms, showing that the kinematic accelerations are very fast. Furthermore, using the same calculations as above and plugging in an additional $0.06g$, the error at the moment when the kinematic acceleration occurs yields:

$$[\alpha \quad \beta \quad \gamma] = [-2.47 \quad 2.46 \quad 0.04]^o$$

$$\|[\alpha \quad \beta \quad \gamma]\| = 3.5^o$$

Consequently, the additional kinematic acceleration during the motion event (occurring at the start and end for a duration of 5-6 ms of approximately 300-400 ms total motion) can be disregarded, as its impact on the rotation estimates will be negligible.

# Supplementary Material 2 - Boston Children's Hospital MRI Safety Test

**Function of the product:** Motion Correction Research and collection of physiological data

**Research or Clinical:** Research

**On site Test results:** The BIOPAC system consists of a data collecting unit which is not safe to enter Zone IV. This unit is to be kept in Zone IV connected to a laptop which collects the data from the physiological data collectors in Zone IV. This is accomplished by passing monitoring devices which are through the waveguides into Zone IV.

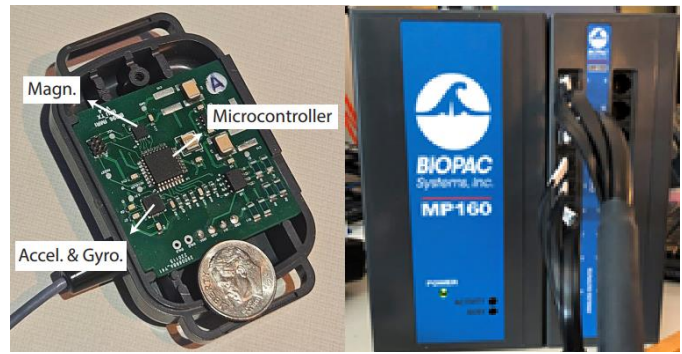

Figure 2. BIOPAC motion sensor device on the left and data collecting unit (digital signal processing box) on the right.

**Static Field Considerations:** The sensor device can be brought to the middle of the bore without any signs of magnetic attraction. The BIOPAC data collecting unit is brought out from storage and set up on the control counter in Zone III. This system is then connected to a laptop which collects data from inside the bore of the scanner.

The twisted copper cables of the sensor device are passed through the waveguides onto the scan table. These cables have a metal component at the end that attaches to the BIOPAC system in Zone III but the end of the fiber cable that is passed through the waveguide does not have any magnetic components.

Both Ferroguard and hand-held ferromagnetic detectors did not detect any ferromagnetic parts in the sensor assembly.

**RF field considerations:** The copper cables were tested for potential RF heating. There was no sign of heating when proper procedure was followed. Cable was directed straight out of the bore free from any looping. The fast brain protocol was performed with the sensor on a phantom. The temperature was monitored and found to increase 0.3 degree Celsius. The cable was significantly insulated.

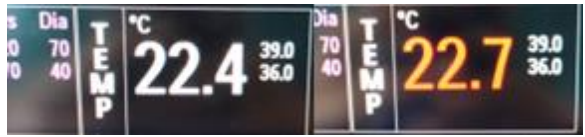

Figure 3. Temperature of the sensor as measured by a temperature probe before and after scanning.

**Imaging considerations:** No artifacts were displayed on the images with examples from the Ax T2 Flair sequence.

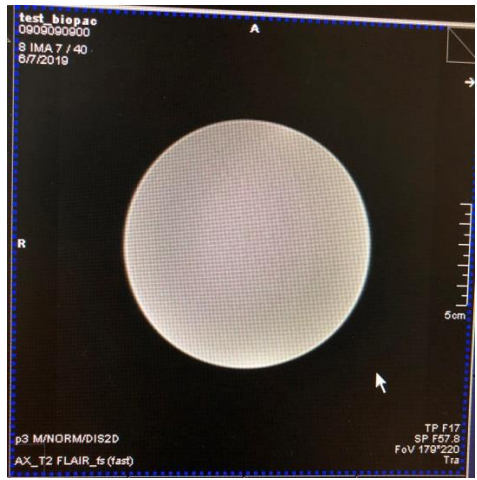

Figure 4. Imaging result with the sensor on the phantom for the Ax T2 Flair sequence. The sensor is not visible to the MRI and does not introduce artifacts to the image.

**Testing group**: Kristina Pelkola, Chief MRI Research Technologist
Onur Afacan, PhD, Manager of the Research Imaging Core.

**Product Classification:** Device is <u>MR Conditional</u> up to 5 Gauss

**Product label:** MR Conditional label applied to device by Boston Children's testing group (listed above). Device is a prototype and has not undergone manufacturer testing.

**Recommendations**: We found the device: <u>MRI Conditional</u>. The sensor device is <u>MRI safe</u>, whereas the data collecting unit <u>must remain outside</u> the 5 Gauss line and should be tethered to prevent any possibility inadvertently being brought closer to the bore of the magnet.